# Phonons and thermal transport in Si/SiO$_2$ multishell nanotubes: Atomistic study

Calina Isacova[1], Alexandr Cocemasov[1], Denis L. Nika[1,*] and Vladimir M. Fomin[1,2,3]

[1]*E. Pokatilov Laboratory of Physics and Engineering of Nanomaterials, Department of Physics and Engineering, Moldova State University, Chisinau MD-2009, Republic of Moldova.*

[2]*Institute for Integrative Nanosciences (IIN), Leibniz Institute for Solid State and Material Research (IFW) Dresden, Dresden D-01069, Germany.*

[3]*National Research Nuclear University MEPhI (Moscow Engineering Physics Institute), 115409 Moscow, Russia.*

*Correspondence: **dlnika@yahoo.com**

**Abstract:** Thermal transport in the Si/SiO$_2$ multishell nanotubes is investigated theoretically. The phonon energy spectra are obtained using the atomistic Lattice Dynamics approach. Thermal conductivity is calculated using the Boltzmann transport equation within the relaxation time approximation. Redistribution of the vibrational spectra in multishell nanotubes leads to a decrease of the phonon group velocity and the thermal conductivity as compared to homogeneous Si nanowires. Phonon scattering on the Si/SiO$_2$ interfaces is another key factor of strong reduction of the thermal conductivity in these structures (down to 0.2 W/mK at room temperature). We demonstrate that phonon thermal transport in Si/SiO$_2$ nanotubes can be efficiently suppressed by a proper choice of nanotube's geometrical parameters: lateral cross-section, thickness and number of shells.

# 1. Introduction

Rapid miniaturization of electronic devices and increasing power consumption require an efficient heat management at nanoscale [1,2]. Thermal properties of nanostructures and different ways of their optimization have been widely investigated both experimentally and theoretically [3–8]. Phonon engineering, i.e. targeted modification of phonon modes in nanostructures to enhance their thermal and/or electrical properties, manifests itself as a powerful tool for the optimization of nanoscale thermal transport [5,9,10]. Nanomaterials with high thermal conductivity, such as graphene, are promising candidates as heat spreaders and interconnects [11–14], while nanomaterials with low thermal conductivity and high electrical conductivity can be used for thermoelectric applications. The efficiency of the thermoelectric energy conversion, figure of merit $ZT$, is directly proportional to the electrical conductivity and inversely proportional to the total thermal conductivity: $ZT = S^2 \sigma T/(\kappa_{ph} + \kappa_{el})$, where $S$ is the Seebeck coefficient, $\sigma$ is the electrical conductivity, $T$ is the absolute temperature, $\kappa_{ph}$ and $\kappa_{el}$ are the phonon and electron thermal conductivities, respectively. Acoustic phonons are the main heat carriers in bulk semiconductors at room temperature (RT) and above. Strong spatial confinement of acoustic phonons in nanostructures significantly changes phonon energy spectra in comparison with the bulk case, resulting in a decrease of phonon group velocities [15–21]. The latter, in combination with an enhancement of phonon boundary scattering, stipulates a reduction of the lattice thermal conductivity in nanostructures as compared with bulk materials [9,15,16,18,20]. A significant reduction of the thermal conductivity leads to an increase of $ZT$, e.g., in $Bi_2Te_3$ quantum-well structures $ZT$ increases up to 13 times in comparison with the bulk value [22]. The silicon-based nanostructures are also prospective for thermoelectric applications despite the fact that bulk silicon is a poor thermoelectric with $ZT \sim 0.01$ at RT [23]. Hochbaum et al. [24] and Boukai et al. [25] demonstrated that significant increase of ZT occurs in Si nanowires due to suppression of phonon transport. Strong reduction of lattice thermal conductivity in cross-section modulated Si nanowires [26–28] may also lead to improvement of their thermoelectric efficiency as compared with bulk Si. A reduction of thermal conductivity was also reported for SiGe nanocomposite materials [3]. It was shown that rise of the Seebeck coefficient is stronger than a possible increase of the electrical resistivity.

Efficient engineering of the acoustic phonon energy spectrum is carried out in self-rolled micro- and nanoarchitectures [29]. The strain-driven roll-up procedure is a powerful high-tech instrument for fabrication of multi-layer micro- and nano-superlattices and their arrays [30–32]. The acoustic phonon dispersion in multishell tubular nanostructures was analyzed within the framework of elastodynamics [33]. It was shown that the number of shells is an important control parameter of the phonon dispersion together with the structure dimensions and acoustic impedance mismatch between the shells. An increase of the number of shells was shown to lead to an appreciable decrease of the average and root-mean-square phonon group velocities. A strong reduction of lattice thermal conductivity in low-dimensional nanostructures (nanowires, thin films and superlattices) as compared with corresponding bulk materials justified diverse proposals to use them for thermoelectric and thermal insulating applications [15,20,24–28].

In the present paper, we employ an atomistic calculation to tackle the phonons and thermal transport in Si/SiO$_2$ multishell nanotubes (NTs). We demonstrate that phonon thermal transport in Si/SiO$_2$ nanotubes can be efficiently suppressed by a proper choice of nanotube's geometrical parameters: lateral cross-section, thickness and number of shells.

The rest of the paper is organized as follows. In Section 2, we describe our theoretical model employed for calculations of phonon modes and the lattice thermal conductivity in Si/SiO$_2$ multishell NTs. Discussions of phonon modes and thermal transport in Si/SiO$_2$ NTs are provided in Section 3. Conclusions are given in Section 4.

## 2. Theoretical model of thermal conductivity in Si/SiO$_2$ multishell nanotubes

We study rectangular multishell nanotubes formed from alternating layers of silicon and silicon dioxide. The number of Si/SiO$_2$ bilayer shells is varied. A scheme of a multishell NT is shown in Figure 1. The external surface of the nanotube is assumed to be free [8,17,19]. The *X* and *Y* axes of the Cartesian coordinate system are located in the cross-sectional plane of the NT and are parallel to its sides, while the *Z* axis is directed along the NT axis. We assume that the NT is infinite along the *Z* axis. The thicknesses of the shells are denoted $d_{x,Si}$ and $d_{y,Si}$ for silicon ($d_{x,SiO_2}$ and $d_{y,SiO_2}$ for silicon dioxide), while the cavity dimensions are $d_{x,cavity}$ and $d_{y,cavity}$. The number of Si/SiO$_2$ bilayer shells is denoted by *N*.

The phonon energy spectra in the Si/SiO$_2$ multishell NTs and Si NWs are calculated using an atomistic Face-Centered Cubic Cell (FCC) model within the Lattice Dynamics approach [8]. In the FCC model, the diamond-type crystal lattice, consisting of two shifted face-centered cubic Bravais sublattices, is replaced with one face-centered cubic lattice with all atoms possessing a doubled mass. The equations of motion for atoms in the harmonic approximation is [26]:

$$m_i\omega^2 u_\alpha(\vec{r}_i,\vec{q}) = \sum_{\beta=x,y,z;\,\vec{r}_j} D_{\alpha\beta}(\vec{r}_i,\vec{r}_j)u_\beta(\vec{r}_j,\vec{q}),\ \alpha = x,y,z, \qquad (1)$$

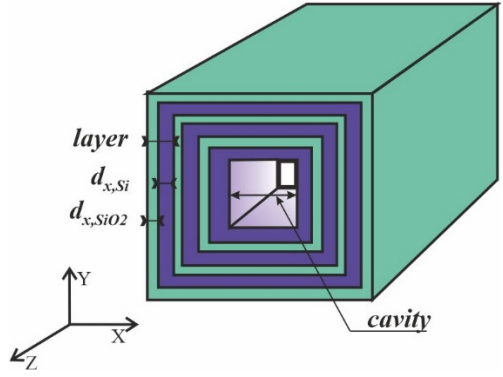

**Figure 1.** Scheme of a Si/SiO$_2$ rectangular multishell NT

Here, $i$ enumerates atoms in a NT translation period (two cross-sectional atomic planes in case of FCC model), $m_i$ is the mass of the $i$-th atom, $\vec{r}_i$ and $\vec{r}_j$ are the radius vectors of the $i$-th and $j$-th atoms, respectively, $\vec{q}$ is the phonon wavevector, $\omega$ is the sought phonon frequency, $u_\alpha(\vec{r}_i, \vec{q})$ is a component of the displacement vector for the $i$-th atom and the dynamic matrix $D_{\alpha\beta}(\vec{r}_i, \vec{r}_j)$ is

$$D_{\alpha\beta}(\vec{r}_i, \vec{r}_j) = \Phi_{\alpha\beta}(\vec{r}_i, \vec{r}_j)/\sqrt{m_i m_j}, \qquad (2)$$

where $m_j$ is the mass of the $j$-th atom and $\Phi_{\alpha\beta}(\vec{r}_i, \vec{r}_j)$ is the matrix of force constants. In the FCC model, the summation in Eq. (1) is performed over all the nearest and second-nearest atoms of the $i$-th atom: 12 nearest atoms at $\vec{r}_j = \vec{r}_i + \vec{h}_j^I$ ($j$=1,…,12) and 6 second-nearest atoms at $\vec{r}_j = \vec{r}_i + \vec{h}_j^{II}$ ($j$=1,…,6) [8,26]. The components of the vectors $\vec{h}_j^I$ and $\vec{h}_j^{II}$ are presented in Table I of Ref. [26]. The force constant matrix $\Phi_{\alpha\beta}(\vec{r}_i, \vec{r}_j)$ used in our calculations is taken from Ref. [8]:

$$\Phi_{\alpha\beta}^{(1)}(\vec{h}^{(k)}, \vec{r}_k, \vec{r}'_k) = \frac{-\kappa_1(\vec{r}_k, \vec{r}'_k) h_\alpha^{(k)} h_\beta^{(k)}}{|\vec{h}^{(k)}|^2}, \qquad (3)$$

$$\Phi_{\alpha\beta}^{(2)}(\vec{h}^{(k)}, \vec{r}_k, \vec{r}'_k) = \begin{pmatrix} \kappa_2(\vec{r}_k, \vec{r}'_k) & 0 & 0 \\ 0 & \kappa_3(\vec{r}_k, \vec{r}'_k) & 0 \\ 0 & 0 & \kappa_3(\vec{r}_k, \vec{r}'_k) \end{pmatrix} \quad \text{for atoms with coordinates} \quad (\pm a, 0, 0),$$

$$\Phi_{\alpha\beta}^{(2)}(\vec{h}^{(k)}, \vec{r}_k, \vec{r}'_k) = \begin{pmatrix} \kappa_3(\vec{r}_k, \vec{r}'_k) & 0 & 0 \\ 0 & \kappa_2(\vec{r}_k, \vec{r}'_k) & 0 \\ 0 & 0 & \kappa_3(\vec{r}_k, \vec{r}'_k) \end{pmatrix} \quad \text{for atoms with coordinates} \quad (0, \pm a, 0), \quad (4)$$

$$\Phi_{\alpha\beta}^{(2)}(\vec{h}^{(k)}, \vec{r}_k, \vec{r}'_k) = \begin{pmatrix} \kappa_3(\vec{r}_k, \vec{r}'_k) & 0 & 0 \\ 0 & \kappa_3(\vec{r}_k, \vec{r}'_k) & 0 \\ 0 & 0 & \kappa_2(\vec{r}_k, \vec{r}'_k) \end{pmatrix} \text{ for atoms with coordinates } (0, 0, \pm a).$$

This matrix depends on 3 independent force constants $\kappa_1$, $\kappa_2$ and $\kappa_3$, which can be expressed through the elastic moduli $c_{11}$, $c_{12}$ and $c_{44}$ of a bulk cubic crystal [8]: $\kappa_1 = \frac{a(c_{12}+c_{44})}{2}$, $\kappa_2 = \frac{a(c_{11}-c_{12}-c_{44})}{4}$, $\kappa_3 = \frac{a(c_{44}-c_{12})}{8}$, where $a$ is the lattice constant.

For calculation of the thermal conductivity of Si/SiO$_2$ multishell NTs and Si NWs, we use the following expression, which was derived from the Boltzmann transport equation within the relaxation time approximation [6,20,34,35] taking into account quasi one-dimensional density of phonon states:

$$\kappa_{ph} = \frac{1}{2\pi k_B T^2 S_{NT}} \sum_s \int_0^{q_{z\,max}} [\hbar\omega_s(q_z)v_{z,s}(q_z)]^2 \tau_{tot,s}(q_z) \frac{\exp\left(\frac{\hbar\omega_s(q_z)}{k_B T}\right)}{\left[\exp\left(\frac{\hbar\omega_s(q_z)}{k_B T}\right)-1\right]^2} dq_z, \quad (5)$$

In. Eq. (5), the summation is performed over all phonon branches $s = 1, \ldots, N_b$, $S_{NT}$ is the NT cross-sectional area, $\omega_s$ is the phonon frequency, $v_{z,s}$ is the Z-th component of the phonon group velocity, $\tau_{tot,s}$ is the total phonon relaxation time, $k_B$ is the Boltzmann constant, $\hbar$ is the Planck constant and $T$ is the temperature.

The total phonon relaxation rate was estimated according to the Matthiessen's rule:

$$\tau_{tot,s}^{-1}(q_z) = \tau_{U,s}^{-1}(q_z) + \tau_{imp,s}^{-1}(q_z) + \tau_{x,s}^{-1}(q_z). \quad (6)$$

Here, $\tau_{U,s}$ is the relaxation time for the Umklapp scattering: $\tau_{U,s}^{-1}(q_z) = B(\omega_s(q_z))^2 T \exp(-C/T)$ [20] and $\tau_{imp,s}$ is the relaxation time for the phonon-impurity scattering: $\tau_{imp,s}^{-1}(q_z) = A[\omega_s(q_z)]^4$ [8,20]. We performed two calculations of the thermal conductivity using different values of the third term in Eq. (6): with the „effective scattering rate" of phonons $\tau_{SiO_2}^{-1}$ due to diffusion in amorphous SiO$_2$ [32]:

$$\frac{1}{\tau_x} = \frac{1}{\tau_{SiO_2,s}(q_z)} = v_{z,s}^2(q) \frac{3\omega_s(q)}{a_{b.l.}^2 F\langle\omega\rangle^2}, \quad (7)$$

or, alternatively, with the relaxation rate for the boundary scattering $\tau_B^{-1}$:

$$\frac{1}{\tau_x} = \frac{1}{\tau_{B,s}(q_z)} = \frac{1-p}{1+p} \frac{|v_{z,s}(q_z)|}{2} N \left(\frac{1}{d_{x,Si}} + \frac{1}{d_{y,Si}} + \frac{1}{d_{x,SiO_2}} + \frac{1}{d_{y,SiO_2}}\right), \quad (8)$$

where $p$ is the specularity parameter, $a_{b.l.} = 0.235$ nm is the SiO$_2$ bond length, $\langle\omega\rangle$ is the mean vibrational frequency (the mean vibrational energy $\hbar\langle\omega\rangle = 34\ meV$ is taken from Ref. [32]), $F = 0.33$ according to Ref. [36].

**3. Results and Discussion**

We calculated the phonon energy spectra in Si/SiO$_2$ multishell NTs and Si NWs by solving Eq. (1) numerically. Our calculations were performed for all $q_z$ in the interval $(0, \frac{\pi}{a})$. In 1D case, the partial phonon density of states (DOS) per unit length in real space can be found from the relation: $g_s(\omega)d\omega = \frac{1}{2\pi}dq_{z,s}$. Hence, the total phonon DOS per unit length in real space

$$g(\omega) = \sum_{s(\omega)} g_s(\omega) = \sum_{s(\omega)} \frac{1}{2\pi v_{z,s}}, \tag{9}$$

where summation is performed over all phonon modes $s(\omega)$ with frequency $\omega$.

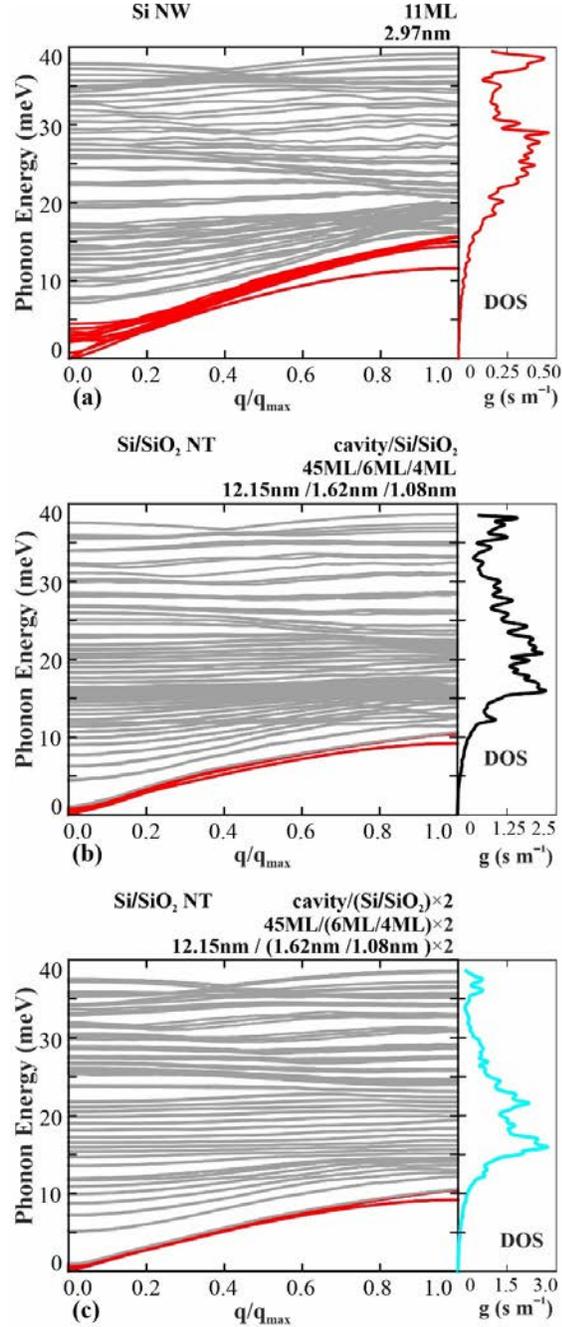

**Figure 2.** Phonon dispersions and phonon density of states in the Si NW with cross-section 11 ML x 11 ML (panel (a)); in the Si/SiO$_2$ NTs with the cavity cross-section 45 ML x 45 ML and one Si/SiO$_2$ bilayer shell (panel b) and two Si/SiO$_2$ bilayer shells (panel c). The thickness of silicon and silica layer in the Si/SiO$_2$ bilayer is 6 ML and 4 ML, correspondingly.

The phonon energy spectra and DOS (in arbitrary units) for Si NW and Si/SiO$_2$ multishell NTs are shown in Figure 2. The Si NW has dimensions 11 ML × 11 ML (2.97 nm × 2.97 nm; 1 ML = 0.27 nm). The geometric parameters of the Si/SiO$_2$ NTs are as follows: the cavity cross-section is 45 ML × 45 ML (12.15 nm × 12.15 nm), the thickness of the Si layer is 6 ML (1.62 nm) and the thickness of the SiO$_2$ layer is 4 ML (1,08 nm). Panel (a) represents results for the Si NW, panel (b) – for the NT with a single Si/SiO$_2$ bilayer shell ($N=1$) and panel (c) – for the NT with two Si/SiO$_2$ bilayer shells ($N=2$). Similarly to rectangular nanowires [19,37], phonon modes in rectangular NTs can be classified into four types according to the spatial symmetry of the displacement vector components: Dilatational, Flexural$_1$, Flexural$_2$ and Shear. For the Si NW, we show phonon branches with numbers $s$ = 1-4, 11, 17, …, 47, 51, 76, … , 301, 320 for Shear polarization; $s$ = 1-4, 11, 17, … , 47, 51, 76, …, 326, 342 for Dilatational polarization and $s$ = 1-4, 11, 17, …, 47, 51, 76, …, 326, 331 for Flexural$_1$ and Flexural$_2$ polarizations. For the Si/SiO$_2$ NT with $N = 1$, we present phonon branches with $s$ =1-5, 30, 55, 480, 500, 550, …, 1000, 1075, … , 1600, 1661 for Dilatational polarization; $s$ = 1-5, 30, 55, 480, 500, 550, …, 1000, 1075, … , 1600, 1640 for Shear polarization and $s$ = 1-5, 30, 55, 480, 500, 550, …, 1000, 1075, … , 1600, 1650 for Flexural$_1$ and Flexural$_2$ polarizations. For the Si/SiO$_2$ NT with $N = 2$, the following phonon branches are depicted: $s$ = 1-5, 80, 155, 230, 305, 380, 455, 500, 675, 850, … , 2950, 3000, 3100 …, 3900, 3920 for Dilatational polarization; $s$ = 1-5, 80, 155, 230, 305, 380, 455, 500, 675, 850, … , 2950, 3000, 3100 …, 3800, 3880 for Shear polarization and $s$ = 1-5, 80, 155, 230, 305, 380, 455, 500, 675, 850, … , 2950, 3000, 3100, … , 3900 for Flexural$_1$ and Flexural$_2$ polarizations.

The number of confined phonon branches in the NTs ($N_b$ = 6600 for a NT with single shell and $N_b$ = 15600 for a NT with double shell) is substantially larger than that in a NW ($N_b$ =1323). The slope of the lowest phonon branches in NTs is smaller than that in a NW due to acoustic mismatch between silicon and silicon dioxide. A great number of phonon modes in the NTs with energy $\hbar\omega > 10$ meV are nearly dispersionless and possess group velocities close to zero. As compared to a NW, the DOS maximum in NTs is shifted toward the lower energy interval, where the drop of the average phonon group velocity is more significant (see Figure 3). The DOS maximum in a NT with two bilayer shells is more prominent due to a larger number of phonon modes concentrated in SiO$_2$, which possess a smaller maximal phonon energy than Si.

The effect of phonon deceleration in NTs with different numbers of Si/SiO$_2$ bilayer shells ($N=1$, 2, 3, 4) is illustrated in Figure 3, where the average phonon group velocity $\langle v \rangle(\omega) = \sum_{s(\omega)}(v_{z,s} g_s(\omega)) / \sum_{s(\omega)} g_s(\omega)$ is shown as a function of the phonon energy for a Si NW with cross-section $N_x \times N_y = 11\text{ML} \times 11\text{ML}$ and different Si/SiO$_2$ NTs with the following geometrical parameters: $N_{cavity,x} \times N_{cavity,y} = 45\text{ML} \times 45\text{ML}$, $N_{Si,x} = N_{Si,y} = 6\text{ML}$, $N_{SiO_2,x} = N_{SiO_2,y} = 4\text{ML}$ (panel (a)) and $N_{cavity,x} \times N_{cavity,y} = 5\text{ML} \times 5\text{ML}$, $N_{Si,x} = N_{Si,y} = 6\text{ML}$, $N_{SiO_2,x} = N_{SiO_2,y} = 4\text{ML}$ (panel (b)). The phonon group velocity in NTs is smaller than that in a NW over a wide energy range (0 to 30 meV). The reduction of the phonon group velocities in NTs is explained by an acoustic mismatch between Si and SiO$_2$, a stronger phonon confinement, and a spectral redistribution of the phonon DOS in NTs. In the energy range from 30 to 37 meV, the phonon group velocities vary near the same low values for all considered NTs. Increase of the number of shells slightly reinforces the drop of phonon group velocity.

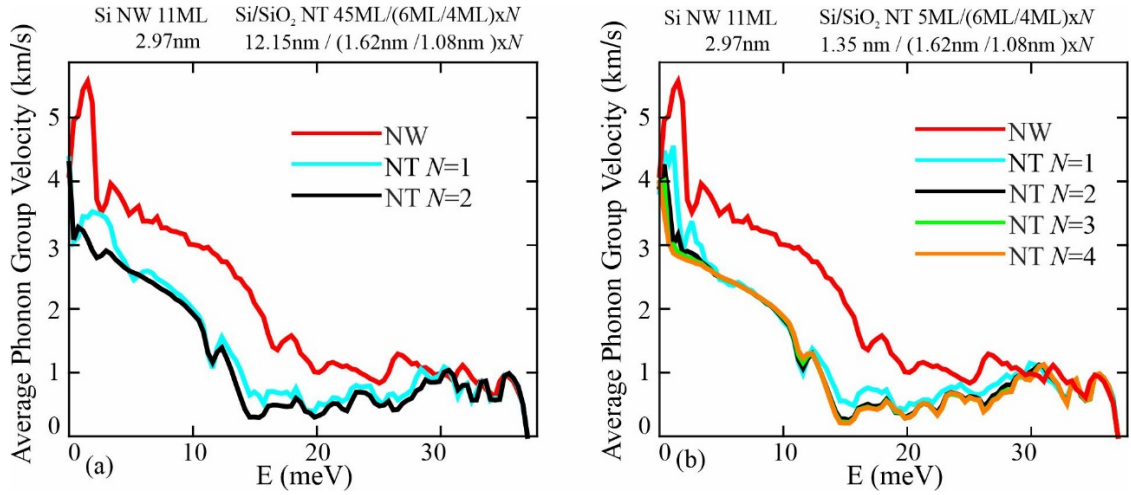

**Figure 3.** Average phonon group velocities as a function of phonon energies in Si/SiO$_2$ NTs with the cavity cross-section 45 ML x 45 ML (a) and 5 ML x 5 ML (b) and different numbers of Si/SiO$_2$ bilayer shells, formed by silicon layer with thickness 6 ML and silica layer with thickness 4 ML. Results for Si NW with cross-section 11 ML x 11 ML are also shown for comparison.

A comparison of temperature dependencies of thermal conductivity (TC) in a Si NW and Si/SiO$_2$ NTs is provided in Figure 4. The geometrical parameters of a NW and NTs are the same as in Figure 3. The thermal conductivity curves presented in panels (b) and (e) are calculated considering diffusion transport in SiO$_2$ (see Eq. 7), while those in panels (a) and (d) – considering boundary scattering (see Eq. 8) with the specularity parameter $p = 0.6$. The values of the thermal conductivity in the NTs (even for the NTs with a greater number of phonon modes) are lower than the thermal conductivity in the NW in the whole temperature range. This decrease of TC in the NTs is due to modification of phonon energy spectra in NTs leading to phonon deceleration, as well as enhancement of the phonon scattering at interfaces and phonon diffusion in SiO$_2$ layers.

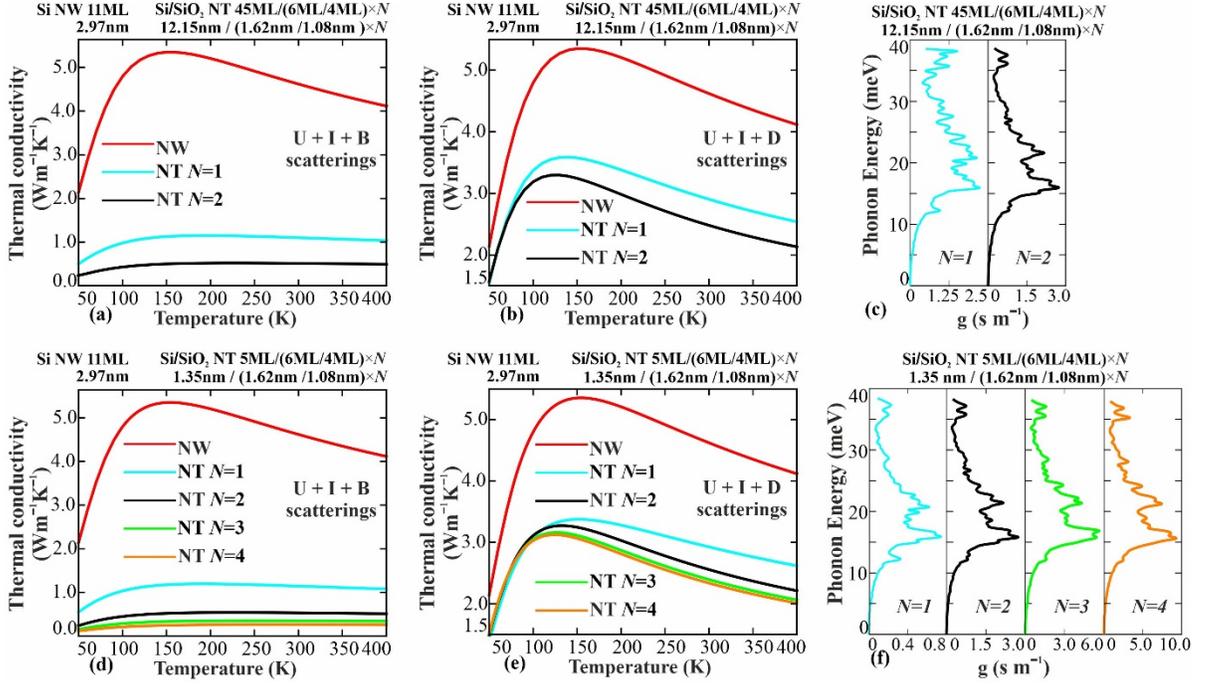

**Figure 4.** Temperature dependence of the TC (a, b, d, e) and phonon DOS (c, f) in Si/SiO$_2$ NTs. The TC is calculated considering the phonon Umklapp and boundary scatterings (a, d) and the phonon Umklapp and "effective" diffusive scattering in SiO$_2$ (b, e).

The maxima of the thermal conductivity curves are determined by the interplay between three-phonon Umklapp and boundary or effective diffusion scatterings. At low temperatures, the boundary scattering dominates, and thermal conductivity increases with temperature due to the population of high-energy phonon modes and approaches a maximum value, when $\tau_U \sim \tau_B$. A subsequent rise of temperature leads to an enhancement of the Umklapp scattering and a diminution of the TC. Additional interfaces between the shells effectively scatter phonons, hence the TC decreases with augmentation of the number of shells (see panels (a) and (d)) reaching values as low as 0.2 W/mK at RT for Si/SiO$_2$ NT with smaller cavity and $N$=4. The similar dependence of TC on $N$ was demonstrated experimentally for Si/SiO$_2$ rolled-up nanotubes of 1.9 □m to 3.2 □m radii and 24 nm-thick shell [32]. The latter fact confirms the importance of our theoretical findings for an accurate prediction of thermal transport phenomena in Si/SiO$_2$ nanotubes in a wide range of lateral cross-sections, number and thicknesses of the shells. However, the phonon dispersion and transport properties in multishell nanostructures with sizes of the order of a few hundred nm require a further analysis because they are in the transitional region between the areas of applicability of elastodynamics and the atomistic approaches.

## 5. Conclusions

Phonon and thermal properties of Si/SiO$_2$ multishell nanotubes were investigated within the Lattice Dynamics approach. The TC in the Si/SiO$_2$ NTs is lower than that in the Si NW with the similar lateral dimensions due to the redistribution of phonon energy spectra in NTs and a stronger phonon confinement. A significant number of phonon modes are scattered on Si/SiO$_2$ interfaces, which enhances the influence of phonon boundary scattering on the TC in the NTs under analysis. As a result, the TC decreases with increase of the number of Si/SiO$_2$ shells. Low values of the TC in Si/SiO$_2$ multishell NTs as compared with other low-dimensional nanostructures make advanced semiconductor multishell NTs prospective candidates for thermoelectric applications.

*Author Contributions:* Conceptualization, D.N. and V.F.; software, C.I., A.C. and D.N.; validation, C.I., A.C., D.N. and V.F.; visualization, C.I.; writing—original draft preparation, C.I. and A.C.; writing—review and editing, D.N. and V.F.; funding acquisition and project administration, D.N. and V.F. All authors have read and agreed to the published version of the manuscript.

*Funding:* This research has been supported by the German Research Foundation (DFG) through grant # FO-956/4-1. C.I. is thankful for kind hospitality during her research stay at the IIN, IFW Dresden. D.N., C.I. and A.C. acknowledge the support by the government of the Republic of Moldova within project #20.80009.5007.02. V.F. acknowledges a partial support through the MEPhI Academic Excellence Project (Contract # 02.a03.21.0005).

*Acknowledgments:* The authors are grateful to O. G. Schmidt and G. Li for fruitful discussions.

*Conflicts of Interest:* The authors declare no conflict of interest. The funders had no role in the design of the study; in the collection, analyses, or interpretation of data; in the writing of the manuscript, or in the decision to publish the results.